# Manipulating Acoustic and Plasmonic Modes in Gold Nanostars

Sharmistha Chatterjee[1,2,3], Loredana Ricciardi[2,3], Julia I. Deitz[4,5], Robert E.A. Williams[4], David W. McComb[4,5] and Giuseppe Strangi*[1, 2, 3].

[1]Department of Physics, Case Western Reserve University, 10600 Euclid Avenue, Cleveland, OH 44106, USA.

[2]CNR-NANOTEC Istituto di Nanotecnologia and Department of Physics, University of Calabria, 87036-Rende, Italy.

[3]Fondazione con Il Cuore, via Roma 170, 88811 Ciro' Marina, Italy.

[4]Center for Electron Microscopy and Analysis, The Ohio State University, Columbus, OH 43212, USA.

[5]Department of Material Science and Engineering, The Ohio State University, Columbus, OH 43210, USA.

**ABSTRACT:** In this contribution experimental evidence of plasmonic edge modes and breathing acoustic modes in gold nanostars (AuNS) is reported. AuNS are synthesized in a surfactant-free, one-step wet-chemistry method. Optical extinction measurements of AuNS confirm the presence of localized surface plasmon resonances (LSPRs), while electron energy-loss spectroscopy (EELS) in the scanning transmission electron microscope (STEM) shows the spatial distribution of LSPRs and reveals the presence of acoustic breathing modes. Plasmonic hot-spots generated at the pinnacle of the sharp spikes, due to the optically active edge dipolar mode, allow significant intensity enhancement of local fields, hot-electron injection, and thus useful for size detection of small protein molecules. The breathing modes observed away from the apices of the nanostars are identified as stimulated dark modes - they have an acoustic nature - and likely originate from the confinement of the surface plasmon by the geometrical boundaries of a nanostructure. The presence of both types of modes is verified by numerical simulations. Both these modes offer the possibility to design nanoplasmonic antenna based on AuNS, which can provide information both on mass and polarizability of biomolecules using a two-step molecular detection process.

1. INTRODUCTION

Label-free detection of protein molecule in their natural state at ultralow concentration is considered as holy-grail in biomedical research.[1] But, because of the acutely small size (< 3 nm) of single protein molecules, their detection becomes exceptionally challenging.[2] One method to deal with this problem is to use the well-known localized surface plasmon effect of noble metal nanoparticles (NPs) which has been used for a wide variety of applications[3-7] including sensing,[8-10] imaging,[11] surface enhanced Raman spectroscopy (SERS),[12-13] quantum technologies and miniaturized photonic circuits.[14-15] Localized surface plasmon resonance (LSPR) of noble metal NPs can be tuned by changing their size, shape, material and the surrounding dielectric matrix.[16-17] NPs with sharp corners like nanotriangle,[18] nanocubes,[19-20] nanorods,[21-22] nanostars,[23-26] or octahedral nanoparticles,[27-28] are capable to confine light over ultra-small regions tightly because of the lightning rod effect along with the plasmonic resonance effect, resulting a higher electromagnetic energy concentration and thus a higher electric field intensity at their hot-spots compared to the non-spiky NPs.[29-31] Thus spiky gold nanoparticles are ideal for plasmonic sensing because of their biocompatibility, tunability and the large field enhancement at their hotspot. A small change in the surrounding dielectric media after the adsorption of protein molecule at the hot-spot results in shift of the NPs LSPR and thus helps to detect the presence of biomolecules. This opens new opportunities for design of next generation nano-devices for sensing applications.

Surface plasmons confined within the geometrical boundaries of flat nanoparticles gives rise to radially symmetric plasmonic breathing modes.[32-33] A flat metal NP not only has edge modes[34] (dipolar, quadrupolar and higher order multipolar modes) because of LSPR, but also supports film modes. Breathing modes are dark modes that cannot be detected by photons as their net dipole moment is zero. However such modes can be detected by inelastic electron scattering in electron energy-loss spectroscopy as the electron wavelength is much smaller than the nanoparticle size.[35-36] The breathing modes are very important for near field coupling effects as they have a very high optical mode density. Several groups have investigated breathing modes of different metal nanostructures such as nanodisk,[32,35,37] nanoplates,[36] core-shell nanoparticles,[38-39] nanotriangles,[40] nanowires,[41-42] nanosphere-nanodisk trimers,[43] metal oligomers,[44] graphene nanoellipses[45] but breathing modes of AuNSs have never been reported before.

AuNSs, because of having several polarization insensitive hot-spots generated after the interaction of light, at the tip of the spikes randomly distributed over their core, are more advantageous than the other spiked nanoantennas like nanoellipsoides and nanorods [46-51] which have same capability to concentrate light like nanostars. AuNSs which are well known for their biomedical applications due to their low-toxicity, biocompatibility, high tunability and high electric field intensity enhancement at their hot-spots [46-49] have been synthesized using nano-chemistry strategies, including environmentally sensitive, "green" synthesis route [52-58] and surfactant-free route. Here, for our study we have synthesized these highly tunable, stable, efficient AuNSs using a low-cost, simple, one-step (seedless), surfactant-free, high-yield wet chemistry method.[59]

In this contribution we report experimental evidence of both the plasmonic edge modes and breathing acoustic modes in AuNS. Results from optical and electron spectroscopy characterization of these highly stable nanoparticles (stability>5 months in aqueous solution) are reported. Optical characterization provided integrated information regarding the collective behaviour of AuNSs in aqueous suspension while electron energy-loss spectroscopy (EELS), done in the scanning transmission electron microscope (FEI Titan$^3$ G2 STEM) provide local plasmonic response of a single AuNS with high spatial resolution. Several groups have investigated plasmonic nanostructures including AuNS via EELS,[60-74] but this high-resolution, low-loss EELS investigations in AuNS have shown the presence of regular edge plasmon modes along with radial breathing modes, irrespective of the spike length.

This study is also supported by extensive theoretical investigations. The effect of tip displacement in response to excitation of the breathing modes has been calculated using the structural mechanics model of Comsol 5.4 which is based on Finite Element Method. The maximum intensity enhancement of single AuNS antenna has been calculated using the Radio Frequency (RF) module of Comsol 5.4. All these results indicate the possibility of creating a device based on the acousto-plasmonic AuNS antenna which will be useful for two-step clinical diagnostics.

2. **RESULTS AND DISCUSSION**

Gold nanostars are synthesized using one-step (without seed) surfactant-free wet chemistry method as described elsewhere.[59] The method is briefly written in the experimental section. Figure 1(a) shows a typical higher magnification TEM image of a gold nanoparticle while Figure 1(b) shows a lower magnification TEM image where almost all the nanoparticles have at least one spike in their surface, confirming the relatively high yield of the synthesis method. X-ray energy dispersive spectroscopy (XEDS) of the synthesized AuNS confirms the presence of Au in the nanostars. (Figure **S1).**

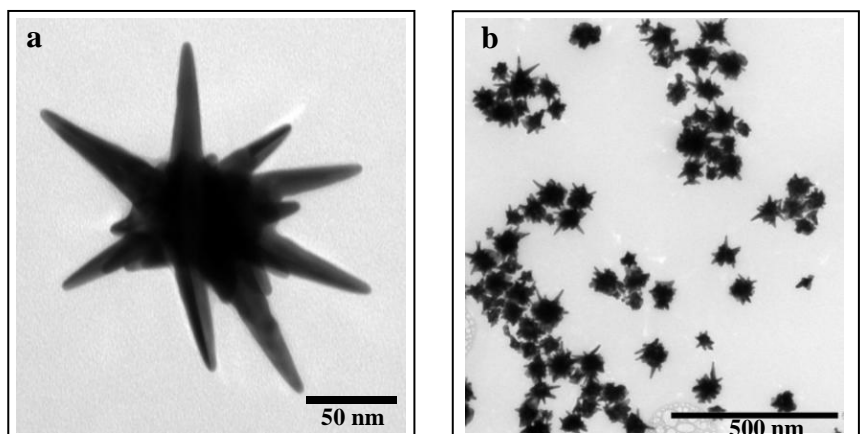

Figure 1. (a) A higher magnification TEM image of a randomly selected gold nanoparticle is shown. (b) A lower magnification TEM image is shown here which is representative of almost all the nanoparticles observed.

Figure 2(a) shows a comparative study between the normalized experimental extinction spectra of the stable AuNS solution collected using a Perkin Elmer Lambda 900 spectrophotometer and the corresponding theoretical investigations. Notably, in the experimental extinction spectra two major modes can be seen. The histogram for spike lengths of synthesized AuNS is shown in Figure 2(b). The average spike length (ASL) of the nanostars was measured and averaged from TEM images of nearly 100 AuNS nanoparticles and found to measure 70 nm while the average diameter of the core measured almost 60 nm. The tip radius of the AuNS spike is varied here from 5 nm to 1 nm depending on the spike length. From the histogram in Figure 2(b) one may observe that the AuNS with an ASL of 70 nm are dominant in solution. AuNS with large spike length (LSL) of 90 nm are the second most dominant type of nanostar in the solution. The histogram found in Figure 2(b), helps support the origin of the two peaks observed in the experimental extinction spectra. Complementary, Finite Element Method (FEM) simulations were performed using Comsol 5.4 for the AuNS with ASL of 70 nm and LSL of 90 nm and having the same core diameter (60 nm), permitting for determination of extinction characteristics. The red and blue curve, shown in Figure 2(a), represent the theoretically calculated normalized extinction spectra for AuNS of spike length 70 nm and 90 nm, the convolution of which gives the resultant theoretical extinction spectra of the AuNS solution. From this figure we can conclude that the resultant theoretical extinction characteristics match well with the experimental one and thus confirming our hypothesis.

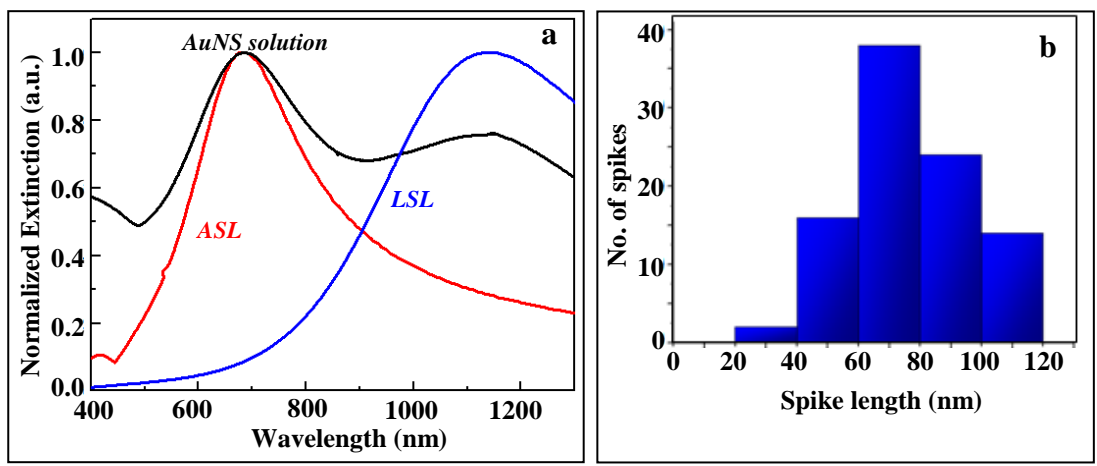

Figure 2. (a) Comparison of normalized experimental extinction spectrum of synthesized AuNS solution and the relevant theoretical investigation done on the extinction property of AuNS of two different spike lengths. Large spike length (LSL) and average spike length (ASL) is taken based on the collected TEM information. (b) Histogram of the spike lengths of synthesized AuNS is shown based on the collected TEM images of nearly 100 NPs.

Although optical and computational studies of AuNS dissolved in aqueous solution provided integrated information about the ensemble behaviour of AuNSs, STEM-EELS measurements revealed localized information about the plasmonic field distribution and resonances for different locations across a single nanostructure. Figure 3 illustrates the EELS analysis for synthesized AuNS with LSL. EELS experiments allowed mapping of plasmon edge modes in AuNS and also the excitation and observation of optically dark radial breathing modes. The STEM-HAADF (High Angle Annular Dark Field) image of an AuNS with LSL which was used for EELS analysis is shown in figure 3(a). The regions from which the EELS spectra were mapped and acquired are indicated by the colorized boxes on the STEM image. Figure 3(b) and 3(c) show the EELS spectra taken at the nanostar core and the nanostar spike. Figure 3(b) shows a peak at 2.2 eV, and it corresponds to the core mode of the AuNS as confirmed by the FEM simulation (**S2**), which shows LSPR at 550 nm (2.2 eV) for a gold core of 60 nm. The strongest mode observed, at the pinnacle of the AuNS spike, was 1.17eV, corresponding with one extinction peak from the UV-Vis-NIR spectra. FEM simulations carried out on similar type of AuNS, having a 88 nm spike length and 60 nm core, support that the LSPR is expected at 1060 nm (1.17 eV) and thereby confirm the dipolar (bright) edge mode nature at the pinnacle (see Figure **S3**).

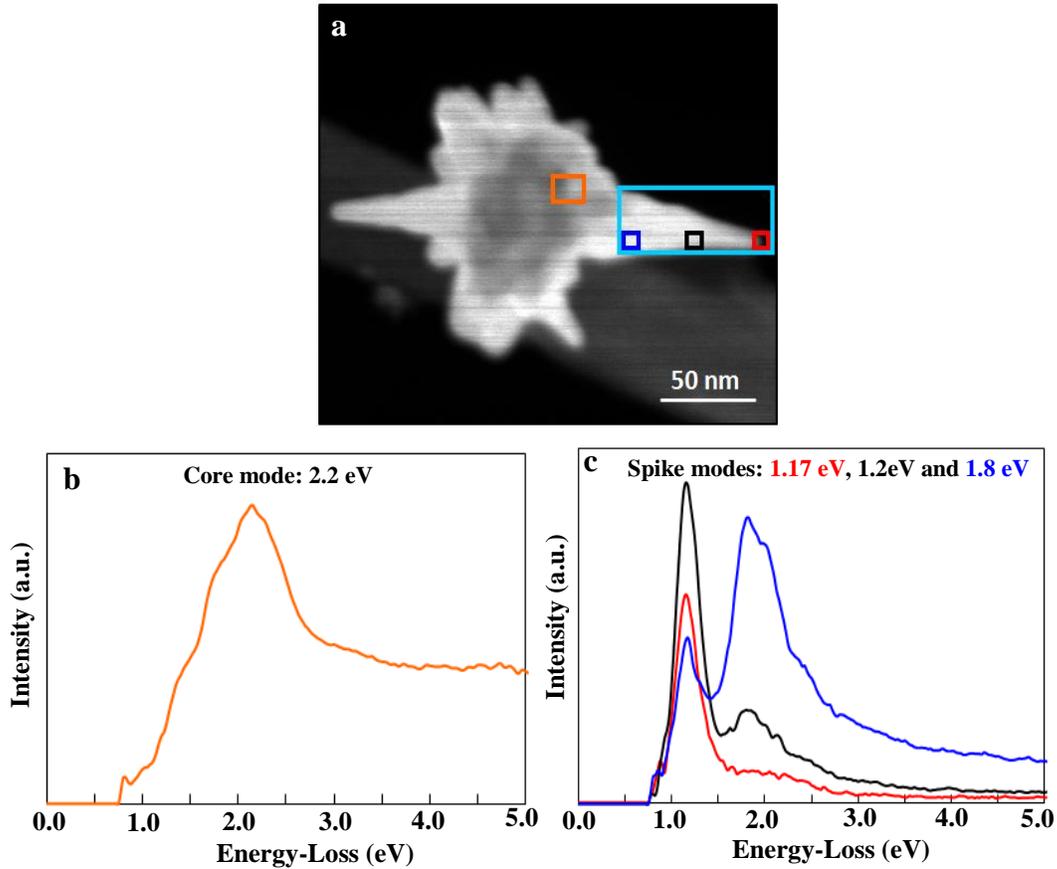

Figure 3. EELS characterization of AuNS of LSL: (a) AuNS with relative areas of investigation indicated by different colored boxes. (b) EELS spectra of AuNS core. (c) EELS spectra of different regions of the AuNS spike.

The EELS intensity maps for 1.2 eV and 1.8 eV, shown in Figure 4(b), (c), confirms the non-plasmonic edge nature observed at the spike of this AuNS.

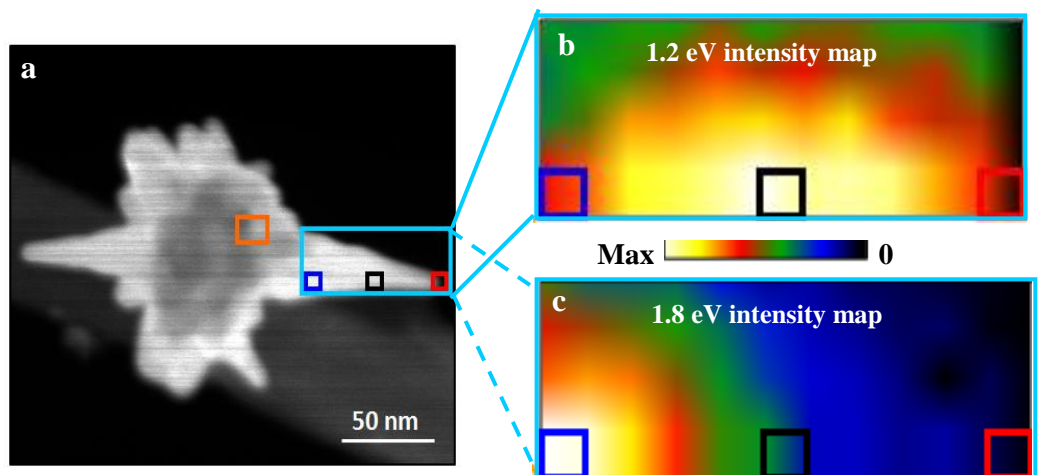

Figure 4. EELS Intensity maps of different dominant modes in spike of AuNS with LSL: (a) AuNS with relative areas of investigation indicated by different colored boxes. (b) EELS Intensity maps of major plasmonic modes, 1.2 eV and 1.8 eV, located at the AuNS spike.

During STEM-EELS mapping, a 1.2 eV mode (Figure 4(b)) was excited and observed in the central region of the spike. This mode shows a very different spatial and intensity distribution with respect to the edge mode. In addition, further away from the pinnacle, a similarly structured mode was observed at 1.8 eV. The comparison of the optical and EELS spectra of several nanostar spikes (small and large spike length) shows that the modes in the body of the spike are dark modes, since they cannot be excited by light and they are not observed in the optical spectra. To further confirm their non-plasmonic edge mode character a computational study was performed using FEM based Comsol 5.4 by analysing the mode shape at each mode frequency. Comsol simulations confirmed that the 1.2 eV mode possess all the features of a radial breathing mode (Figure 6) and it does not present the characteristic spatial distribution of plasmonic edge modes (dipolar, quadrupolar and higher order modes). However, the EELS maps were taken by exciting the entire AuNS with a defocused electron beam; unlike the spectra of Figure 3 taken by spectrum imaging (SI) with a sub-nm electron probe. EELS intensity maps at 1.2 eV and 1.8 eV, show clearly the mode localization away from the AuNS tip.

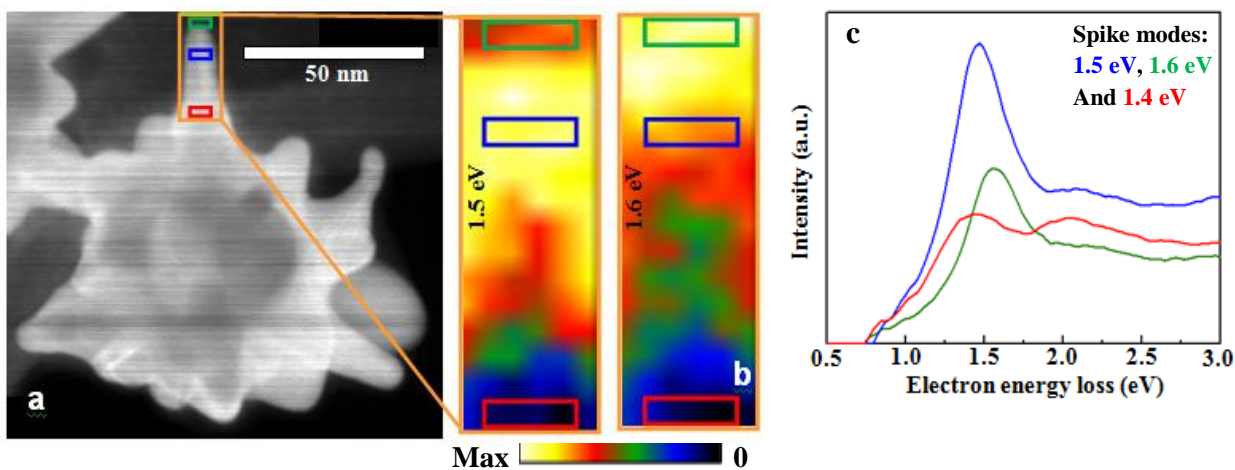

Figure 5. EELS of AuNS of short spike length: (a) AuNS image with relative areas of investigation (coloured boxes). (b) Intensity maps of major plasmonic modes at 1.5 eV and 1.6 eV located at the AuNS spike. (c) EELS spectra of different regions of the AuNS spike. The dominant mode at the pinnacle of the spike is 1.6eV. 1.6 eV EELS intensity map of Figure 4(b) confirms its edge mode nature by observing a maximum intensity at the tip of the spike (green box area). Whereas, the 1.5 eV mode, which is dominant at the body of the spike (blue box region) confirms its non-plasmonic nature.

The presence of radial breathing modes and dipolar edge mode is more evident for short spike than for long spike AuNS, because of its larger interaction volume. The details about the EELS analysis of the AuNS with small spike length (SSL) are shown above (Figure 5) and the relevant numerical results are given in Figure 6.

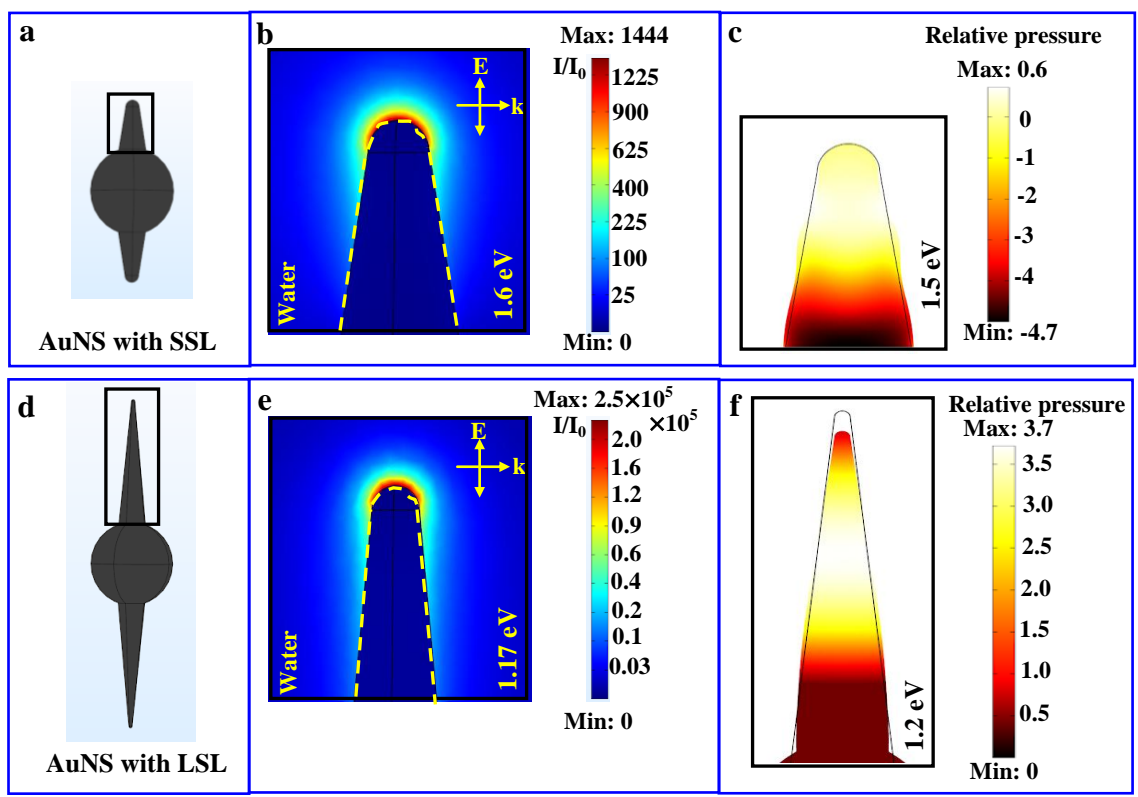

**Figure 6.** (a) Single AuNS with small spike length (SSL) is shown here. The zoomed portion of the black box shows different response at different modes and thus is useful theoretically to predict the nature of the mode (either plasmonic edge mode or radial breathing mode). (b) Behaviour of single AuNS with SSL at 1.6 eV. (c) Behaviour of single AuNS with SSL at 1.5 eV. (d) Single AuNS with LSL is shown. The zoomed portion of the black box shows different response at different modes. (e) Behaviour of single AuNS with LSL at 1.17 eV. (f) Behaviour of single AuNS with LSL at 1.2 eV.

Figure 6 illustrates the numerical study performed for the various observed modes, both edge and non-edge modes, in single AuNS with large spike length (LSL) and small spike length (SSL), when excited by light and electron beams. Figure 6(a) shows an illustration of a plasmonic SSL nanostar antenna determined by the quantitative parameters measured from the STEM images. Figure 6(b) shows the resultant mode distribution at 1.6 eV when excited by light, the mode has its highest intensity at the pinnacle of the AuNS spike, as also observed in the EELS intensity map (Figure 5b). Considering that the skin depth of gold in the wavelength range 400 - 1200 nm is smaller than 5 nm, any plasmonic edge mode is not expected to penetrate in the body of the nanoparticle rather is confined at the metal-dielectric interface. Moreover, extinction studies also confirms the LSPR of this structure at 1.6 eV, thereby, this mode is a dipolar edge mode. Of note, the yellow dashed line in Figure 6(b) represents the nanoparticle edge and does not appear to show evidence of structural modification. The intensity map in Figure 5(b) exhibits an unexpected shape for the 1.5 eV mode, this mode was not observed in the optical extinction spectra, thereby suggesting it forms by a non-plasmonic nature, as also supported by the simulations. The hypothesis that these modes are of acoustic nature was nurtured from multiple observations; the appearance of the highest intensity position away from the metal-dielectric interface, the radially symmetric nature of the mode and the corresponding absence of this mode in the extinction spectra, both for theoretical and experimental studies. To confirm this hypothesis, the mode at 1.5 eV of the SSL AuNS was investigated numerically using a structural mechanics module of Comsol 5.4 (the parameters for the simulation are given in the numerical simulation section).[75-77] Figure 6(c) shows the 1.5 eV mode shape and distribution obtained via Comsol simulations for the SSL AuNS. The map of the relative pressure calculated via the structural mechanics module shows that the mode has the familiar, radial symmetry of a typical breathing mode.[32-34] The black contour line, Figure 6(c), represents the unperturbed morphology of the SSL AuNS, whereas the color map shows the relative pressure distribution at 1.5 eV and the modification of the structure because of the acoustic mode. Similarly, Figure 6(f) reports the simulation obtained via Comsol of the LSL AuNS for the anomalous mode at 1.2 eV, as shown also in Figure 4(b). Notably, the relative pressure is highest in the middle of the spike of the AuNS resulting in a significant contraction of the same structure. The map of the relative pressure, as calculated by structural mechanics, shows that the 1.2eV mode exhibits radial symmetry of a typical breathing mode.[32-34] Again, supporting the acoustic nature of the 1.2 eV mode of the LSL AuNS. Figure 6(e) conversely, shows the results for the analysis of the mode at 1.17 eV of LSL AuNS, as reported in Figure 3. The 1.17 eV mode has its highest intensity at the pinnacle of the AuNS spike, as observed in the EELS spectra (Figure 3c). The 1.17 eV mode is a plasmonic edge mode, since its dipolar nature has been confirmed by calculating the LSPR from the extinction spectra and the electric field mapping using the RF module of Comsol 5.4.

Continued analysis of SI data for breathing modes revealed the 1.5 eV mode is located along the spike where the diameter measured 7.4 nm (Figure 5), whereas for LSL AuNS the breathing mode at 1.2 eV is located along the spike where the diameter is 21.4 nm (Figure 4). For any specific, non-symmetric structures, theoretical calculation of acoustic mode frequency is tedious [78-79] yet Lamb,[80] introduced a very simple relationship between the acoustic breathing mode frequency and the size of the structure. According to Lamb's relationship, the acoustic breathing mode frequency varies inversely with the diameter of the spherically symmetric structures. If the spikes of the AuNS are considered to be a parallel combination of many disks [81], - the Lamb principle is also verified here for the two breathing modes - as the mode frequencies vary inversely with the structure diameter.

The plasmonic edge and breathing modes of AuNSs can be synergistically employed to determine the size and mass of target molecules respectively by combining the reactive sensing principle for plasmonic biosensing [82-83] and a nano-cantilever mechanism. According to the numerical investigation, the field intensity enhancement was found to be approximately $2.5 \times 10^5$ in water media for LSL AuNS at its LSPR (1060 nm). This indicates that the acousto-plasmonic AuNS based nanoantenna – supporting both plasmon edge and breathing modes – can be useful for two-step label free molecular detection at a point-of-care.

## 3. CONCLUSIONS

To summarize, we report the experimental evidence of both plasmonic edge and breathing modes in AuNSs. Here, AuNSs are synthesized via a known simple, one-step, surfactant-free, wet-chemistry method with high yield, and stability and characterized via optical and EELS. Low-loss EELS provided information about the local plasmonic modes of different regions of the nanostars and indicated the presence of optically dark breathing modes. Allowing for the plasmonic edge and breathing modes of the synthesized AuNSs to be harnessed to determine size and mass of adsorbed analyte based on both a plasmon resonance sensing mechanism and the cantilever principle, irrespective of the molecular shape. This study is also supported by the numerical investigations performed by FEM based Comsol 5.4 which confirms the presence of plasmonic edge and breathing modes in different AuNS with different spike lengths. The numerical study predicts that for AuNS nanoplasmonic antenna the field-intensity enhancement factor in the hot-spot region is around $10^5$. This indicates that these efficient acousto-plasmonic AuNS antenna might find applications for a two-step label-free clinical diagnostics.

## 4. EXPERIMENTAL SECTION

**Materials:** Gold (III) Chloride Trihydrate ($HAuCl_4$, $3H_2O$), Silver Nitrate ($AgNO_3$), Ascorbic Acid (AA), Hydrochloric Acid (HCl) (35-37%), Polyvinylpyrrolidone (PVP) were purchased from Sigma Aldrich and used as received without further purification. The water used here was reagent-grade, produced using a Milli-Q SP ultrapure-water purification system.

**Synthesis of the Stabilized Gold Nanostars:** At first, 10 ml of 0.25 mM chloroauric acid ($HAuCl_4$) solution was taken with 10 μl of 1 M HCl solution in a glass vial. After that at room temperature under moderate stirring (700 rpm) 100 μl of 1 mM $AgNO_3$ solution and 50 μl of 100 mM AA solution were added simultaneously. Within few seconds the solution color was appeared to be blue. After 2 minutes from the addition of the AA and $AgNO_3$, 5ml of 2 mM Polyvinylpyrrolidone (PVP) was added and was stirred for another 8 minutes maintaining the same stirring speed. Then the solution was kept for 3 hours at room temperature at rest. Afterwards, one centrifugal wash was done at 4000 rcf for 20 minutes to wash out the extra PVP. After centrifugation keeping the precipitate safely, the liquid containing PVP and the other chemicals washed out and the solution was redispersed in DI water. Finally, the AuNS solution was kept at room temperature for future use. The characterization details of these synthesized AuNS are given below.

**Characterization:** The synthesized nanoparticles are characterized using UV-Vis-NIR spectroscopy and scanning transmission electron microscopy:

**UV-Vis-NIR Spectra:** The UV-Vis-NIR spectra of synthesized gold nanostars solution were taken using a Perkin Elmer Lambda 900 spectrophotometer. The extinction property of the nanostars solution was measured in a wavelength range of 400 nm to 1300 nm.

**STEM, XEDS, and EELS Measurements:** To probe the size and shape of the synthesized nanoparticles STEM (monochromated, image-corrected FEI Titan$^3$ G2 STEM) was used. The electronic structure of the synthesized AuNSs was decoded with the help of low-loss EELS and XEDS helps to determine the gold weight percentage in the solution. Sample preparation has been done one day before the STEM measurements by drop casting and then drying aqueous solution of gold nanoparticles on a standardized holey carbon film supported by a TEM grid. All experiments are done at 60 kV with a high collection angle (~25 mrad) to minimize the influence of Cherenkov radiation in the EELS signal.[84] The convergence angle and the probe size was approximately measured to be 13.2 mrad and 1.3 angstrom respectively. The EELS energy resolution which is equivalent to the full width at half-maximum of the zero-loss peak was approximately 150 meV. To resolve the EELS signals spatially along and across each nano-object spectrum imaging was used. All the EELS data were treated using the Gatan Digital Micrograph software package. No proof of irradiation impairment has been observed in the sample during EEL spectra acquisition. The zero-loss peak for each spectrum was removed using the standard reflected tail method which reflects the tail on the energy-gain side of the spectrum onto the energy-loss side, typically with a pre-defined scaling factor, and subtracts it.[85-88]

**Numerical Simulation:** FEM Simulations are used here to find a correlation between the experimental data and the predicted properties of extinction (Optical Module) and acoustic modes (Structural Mechanics Module) of AuNS. During simulation, data regarding different size and shape of AuNSs spikes and their cores are taken from the STEM studies. The study is carried out to map the electric field for plasmonic modes and structural modification of the nanoantenna for phononic modes. The relative pressure due to the excitation of the breathing modes of AuNS has been calculated using the structural mechanics module of COMSOL 5.4[75-77] where the mode frequency has been taken from the experimental EELS information. Radio Frequency (RF) module of Comsol 5.4 is used to investigate the plasmonic modes. Numerical study has also been done for hybrid AuNS antenna for different size of the nanosphere and gap between each nanoantenna of that heterodimer. Scattering, absorption and extinction cross-section of the AuNS plasmonic antenna are also calculated here. For the optical simulations the definitions in the Comsol material library are used for all the optical and physical properties of the surrounding media. Simulations considering both the air ($\varepsilon = 1.0$) and the water ($\varepsilon = 1.77$) as the homogeneous surrounding media have been done here. During this numerical investigation, the wavelength dependent permittivity of gold is taken from the Johnson and Christy measurements [89] and plane waves are used as the excitation source. Electric field polarization of the incident light is chosen always along the semi major axis of the AuNS spike during extinction and electric field mapping. For the simulation of acoustic modes performed in air the material mechanical properties of Au are defined according to the bulk values. Physics controlled free tetrahedral mesh with extremely fine size for the AuNS and normal size for surrounding media have been chosen for all analysis. For the simulation of acoustic modes, performed in air, the material mechanical properties of Au are defined according to the bulk values (Young's modulus = 79 GPa, Poisson's ratio = 0.4, and density = 19300 kg/m$^3$). The longitudinal speed of the sound in the gold is taken as 3240 m/s.

## ASSOCIATED CONTENT

**Supporting Information**

Additional informations including XEDS results of synthesized nanoparticles along with the FEM simulation results are available in the supporting information.

## AUTHOR INFORMATION

**Corresponding Author**

*E-mail: Gxs284@case.edu.

**Notes**

The authors declare no competing financial interest.
**Corresponding Author**

*E-mail: Gxs284@case.edu.

**Notes**

The authors declare no competing financial interest.

## ACKNOWLEDGEMENTS

We acknowledge support from the Ohio Third Frontier Project 'Research Cluster on Surfaces in Advanced Materials (RC-SAM) at Case Western Reserve University.

# Supporting Information

# Manipulating Acoustic and Plasmonic Modes in Gold Nanostars

*Sharmistha Chatterjee*[1, 2, 3]*, Loredana Ricciardi*[2, 3]*, Julia I. Deitz*[4, 5]*, Robert E.A. Williams*[4]*, David W. McComb*[4, 5]*, Giuseppe Strangi*[\*, 1, 2, 3]*.*

[1]Department of Physics, Case Western Reserve University, 10600 Euclid Avenue, Cleveland, Ohio 44106, USA.

[2]CNR-NANOTEC Istituto di Nanotecnologia and Department of Physics, University of Calabria, 87036-Rende, Italy.

[3]Fondazione con Il Cuore, via Roma 170, 88811 Ciro' Marina, Italy.

[4]Center for Electron Microscopy and Analysis, The Ohio State University, Columbus, Ohio 43212, USA.

[5]Department of Material Science and Engineering, The Ohio State University, Columbus, Ohio 43210, USA.

1. **X-ray Energy Dispersive Spectroscopy (XEDS) of synthesized AuNS:**

X-ray energy dispersive spectroscopy (XEDS) of the synthesized AuNS is shown below. In the spectrum the most intense peak corresponds to carbon due to the carbon substrate used. This spectrum confirms the presence of Au in the synthesized nanostars.

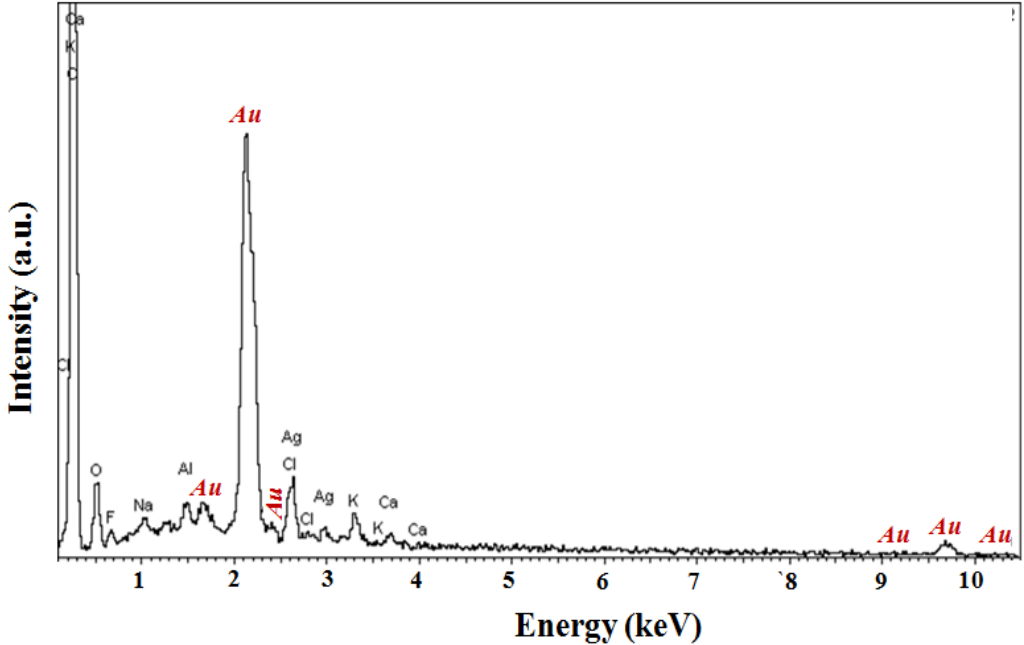

Figure S1. X-ray energy dispersive spectroscopy (XEDS) of the synthesized stable AuNS is shown here. The most intense peak in the XEDS spectrum corresponds to carbon due to the carbon support film. The peak observed at 2.12keV, corresponding to Au, is confirming the presence of Au in the synthesized nanoparticle (from reference [4]).

2. **Numerical Investigation on the Extinction Cross Section of Au Nanosphere of 60 nm size:**

Theoretical extinction cross-section of a 60 nm Au nanosphere is shown here. Radio frequency (RF) module of FEM based comsol 5.4 is used for this calculation which shows that the localized surface plasmon resonance (LSPR) is at 550 nm. According to electron energy loss spectroscopy (EELS) investigation, the mode predominating in the nanostars core was at 2.2 eV (~564 nm). According to TEM study the average core size of synthesized AuNS is 60 nm. So the theoretical evaluation of LSPR at 550 nm of this Au nanosphere is indicating that indeed 2.2 eV mode is the nanostar core mode without significant contribution from the other parts of AuNS. For this study the wavelength dependent permittivity of gold is taken from the Johnson and Christy measurements.[1]

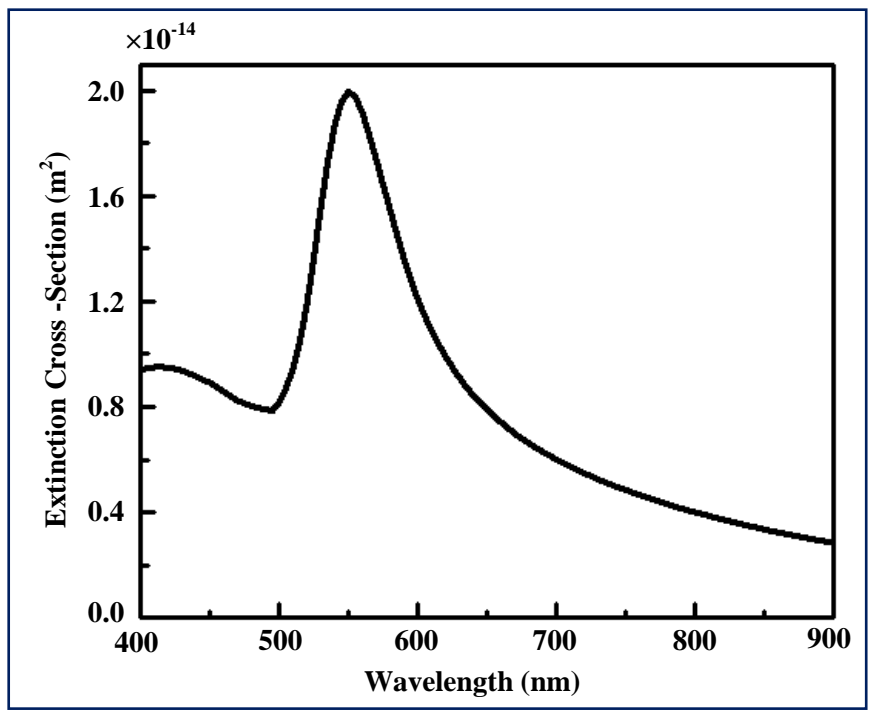

Figure S2. Extinction cross-section of 60 nm Au nanosphere is shown. The theoretical LSPR is seen to occur here at 550 nm.

### 3. Numerical Investigation on the Extinction Property of Single AuNS with LSL:

A theoretical extinction spectrum of AuNS of 60 nm core and 88 nm spike (LSL) is shown here. RF module of Comsol 5.4 is used for this analysis. LSPR of AuNS in water media is found to be at 1060 nm (1.17 eV). Thus, the 1.17 eV mode seen in the EELS study, in AuNS tip area is a dipolar edge mode. Here the incident light wave has polarization along the semi-major axis of the 88 nm spikes of the AuNS. For this study the wavelength dependent permittivity of gold is taken from the Johnson and Christy measurements.[1]

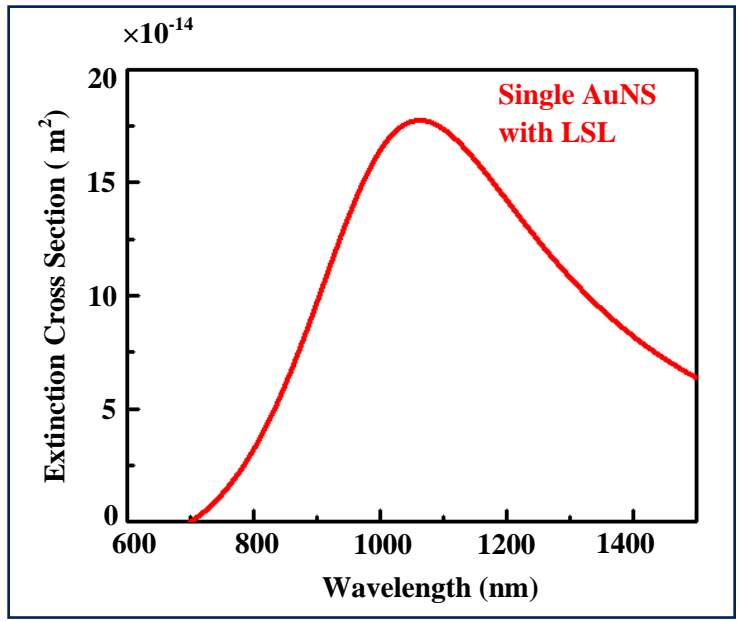

Figure S3. Extinction property of single LSL AuNS in water is shown.